# Nonlinear Kinetic Sunyaev-Zeldovich Effect

Chung-Pei Ma[1,*] and J. N. Fry[2,†]

[1]*Department of Astronomy, University of California at Berkeley, Berkeley, California 94720*
[2]*Department of Physics, University of Florida, Gainesville, Florida 32611-8440*
(Received 19 June 2001; revised manuscript received 19 December 2001; published 13 May 2002)

We derive fully nonlinear expressions for temperature fluctuations from the kinetic Sunyaev-Zeldovich (SZ) effect, the scattering of cosmic microwave background photons off hot electrons in bulk motion. Our result reproduces the Ostriker-Vishniac effect to second order in perturbation theory but contains nonlinear corrections to the electron velocities and densities that were neglected previously. We use the recently developed halo model for nonlinear gravitational clustering to compute the nonlinear kinetic SZ power spectrum, which dominates the primary anisotropy on small angular scales.

*Introduction.*—Scattering of cosmic microwave background (CMB) photons from hot electrons in galaxies and clusters changes the apparent brightness of the CMB, inducing anisotropies in the observed temperature, the Sunyaev-Zeldovich (SZ) effect [1]. The kinetic SZ effect due to scattering off ionized matter in bulk peculiar motion can be separated from the thermal SZ effect due to electron pressure by its different dependence on photon frequency; in particular, the kinetic effect dominates at frequencies about 218 GHz, where the thermal effect vanishes. These foreground contributions are overwhelmed by primary anisotropies for spherical harmonic index $\ell$ smaller than a few thousand but dominate at higher $\ell$ that are being probed by ongoing and upcoming CMB experiments.

The kinetic SZ effect vanishes at leading order in perturbation theory for small inhomogeneities [2,3] (see below). At second order, however, it generates temperature fluctuations on arcminute angular scales, as first pointed out by Ostriker and Vishniac (OV) [4,5]. The OV power spectrum has since been studied in some detail [6–12]. In this Letter we investigate analytically the kinetic SZ effect in the fully nonlinear regime of large density contrasts, obtaining general expressions for the nonlinear momentum power spectrum and presenting results in the halo model. The nonlinear behavior is important because reionized material in the low-redshift universe mostly resides in collapsed, high density objects. Our analytical approach complements recent kinetic SZ studies with hydrodynamical simulations [13] and provides an efficient way to examine the dependence on physical parameters such as the electron density profile, temperature, and reionization redshift.

The fractional temperature distortion due to the kinetic SZ effect along the line-of-sight unit vector $\hat{\gamma}$ is

$$\frac{\Delta T}{T}(\hat{\gamma}) = \int dl\, e^{-\tau} n_e \sigma_T \hat{\gamma} \cdot \boldsymbol{v} = \int dl\, e^{-\tau} \bar{n}_e \sigma_T \hat{\gamma} \cdot \boldsymbol{q}, \quad (1)$$

where $\tau$ is the optical depth from the observer to a scatterer of peculiar velocity $\boldsymbol{v}$ and $\sigma_T$ is the Thomson cross section. The electron density is $n_e = \bar{n}_e(1 + \delta)$, where $\delta$ is the density contrast, $\bar{n}_e = X_e \Omega_b \rho_c (1 + z)^3/m_p$ is the proper mean electron density at redshift $z$, $X_e$ is the ionization fraction, $\Omega_b$ is today's baryon density parameter, $\rho_c$ is the critical density, and $m_p$ is the proton mass. The density-weighted peculiar velocity $\boldsymbol{q}$ and its Fourier transform $\tilde{\boldsymbol{q}}$ are $\boldsymbol{q}(\boldsymbol{x}) = \boldsymbol{v}(1 + \delta)$ and

$$\tilde{\boldsymbol{q}}(\boldsymbol{k}) = \tilde{\boldsymbol{v}}(\boldsymbol{k}) + \int \frac{d^3 k'}{(2\pi)^3} \tilde{\boldsymbol{v}}(\boldsymbol{k}') \tilde{\delta}(\boldsymbol{k} - \boldsymbol{k}'), \quad (2)$$

where $\tilde{\delta}$ and $\tilde{\boldsymbol{v}}$ are the Fourier transforms of $\delta$ and $\boldsymbol{v}$. The angular two-point correlation function of the CMB anisotropy $C(\theta)$ is

$$C(\theta) = \left\langle \frac{\Delta T}{T}(\hat{\boldsymbol{n}}_1) \frac{\Delta T}{T}(\hat{\boldsymbol{n}}_2) \right\rangle = \sum_{\ell=0}^{\infty} \frac{(2\ell + 1)}{4\pi} C_\ell \mathcal{P}_\ell, \quad (3)$$

where $C_\ell$ is the angular power spectrum, and $\mathcal{P}_\ell(\cos\theta) = \mathcal{P}_\ell(\hat{\boldsymbol{n}}_1 \cdot \hat{\boldsymbol{n}}_2)$ are the Legendre polynomials. For $\ell \gg 1$, the $C_\ell$ can be calculated by projecting the 3D power spectrum in the small-angle approximation [7,10,14]. For the kinetic SZ effect, we find

$$C_\ell = \frac{\bar{n}_{e,0}^2 \sigma_T^2}{H_0^2} \int \frac{dx}{x^2} (1 + z)^4 e^{-2\tau} P_{q_\gamma}\!\left(\frac{\ell}{x}, z\right), \quad (4)$$

where $P_{q_\gamma}(k, z)$ is the power spectrum of the line-of-sight component $\hat{\gamma} \cdot \tilde{\boldsymbol{q}}$, $\bar{n}_{e,0}$ is $\bar{n}_e$ at redshift 0, and $x$ denotes the comoving distance; in zero curvature models, $dx = H_0^{-1} dz/\sqrt{\Omega_m (1 + z)^3 + \Omega_\Lambda}$.

The key quantity for us to compute is therefore $P_{q_\gamma}$ in the fully nonlinear theory. Before deriving it, we first note an important feature of the kinetic SZ effect: Because of (near) cancellations in the radial projection in Eq. (4), only modes with orientation $\hat{\boldsymbol{k}}$ perpendicular to $\hat{\gamma}$ contribute to the integrated effect [3,10]. This implies that the density-weighted velocity along the line of sight obeys $\hat{\gamma} \cdot \tilde{\boldsymbol{q}} = \hat{\gamma} \cdot (\tilde{\boldsymbol{q}}_\parallel + \tilde{\boldsymbol{q}}_\perp) \approx \hat{\gamma} \cdot \tilde{\boldsymbol{q}}_\perp$, where we decompose $\tilde{\boldsymbol{q}}$ into a longitudinal component $\tilde{\boldsymbol{q}}_\parallel = \hat{\boldsymbol{k}}(\tilde{\boldsymbol{q}} \cdot \hat{\boldsymbol{k}})$ and a transverse component $\tilde{\boldsymbol{q}}_\perp = \tilde{\boldsymbol{q}} - \hat{\boldsymbol{k}}(\tilde{\boldsymbol{q}} \cdot \hat{\boldsymbol{k}})$ relative to $\hat{\boldsymbol{k}}$. Let $P_{q_\perp}$ and $P_{q_\parallel}$ be the power spectrum for these two components; we then have the simple relation $P_{q_\gamma} \approx \frac{1}{2} P_{q_\perp}$. Although $q_\parallel$ does not contribute to the kinetic SZ effect directly, we will keep track of this component because it has a simple physical meaning: the equation of continuity $\dot{a}\delta + \nabla \cdot [(1 + \delta)\mathbf{v}] = 0$ becomes $\tilde{q}_\parallel(\mathbf{k}, t) = (ia\mathbf{k}/k^2)\dot{\tilde{\delta}}(\mathbf{k}, t)$ in the Fourier domain; $q_\parallel$ is therefore proportional to the time derivative of the density field $\delta$.

Our task now is to compute $P_{q_\perp}$ and $P_{q_\parallel}$. To linear order, we have $\tilde{\mathbf{q}} = \tilde{\mathbf{v}} \propto \hat{\mathbf{k}}$, so $\tilde{\mathbf{q}}_\perp = 0$ and the linear kinetic SZ effect is negligibly small [3]. In the nonlinear regime where $\delta > 1$, the second term of $\tilde{\mathbf{q}}$ in Eq. (2) dominates, so we will focus on this term from this point on. The power spectrum for $\tilde{\mathbf{q}}$ will then be the fourth moment of two $\delta$ and two $\mathbf{v}$ fields. In general, the fourth moment for quantities with zero mean has contributions from second moments of all possible pairs plus an irreducible or connected fourth moment: $\langle ABCD \rangle = \langle AB \rangle \langle CD \rangle + \langle AC \rangle \langle BD \rangle + \langle AD \rangle \langle BC \rangle + \langle ABCD \rangle_c$. Our result therefore contains density and velocity spectra $P_{\delta\delta}$, $P_{vv}$, $P_{\delta v}$, and also the irreducible fourth moment $P_{\delta\delta vv}$. We find the power spectrum for $\tilde{q}_i \tilde{q}_j$ to be

$$P_{qq}^{ij}(\mathbf{k}) = \int \frac{d^3k'}{(2\pi)^3} \frac{d^3k''}{(2\pi)^3} \{(2\pi)^3 \delta_D(\mathbf{k} - \mathbf{k}' - \mathbf{k}'')[\hat{k}'^i \hat{k}'^j P_{vv}(k') P_{\delta\delta}(k'') + \hat{k}'^i \hat{k}''^j P_{\delta v}(k') P_{\delta v}(k'')] + \hat{k}'^i \hat{k}''^j P_{\delta\delta vv}(\mathbf{k} - \mathbf{k}', -\mathbf{k} - \mathbf{k}'', \mathbf{k}', \mathbf{k}'')\}. \tag{5}$$

From $P_{q_\perp} = 2P_{q_\gamma} = 2\hat{\gamma}^i \hat{\gamma}^j P_{qq}^{ij}$ and $P_{q_\parallel} = \hat{k}^i \hat{k}^j P_{qq}^{ij}$, we then obtain the main equation in this Letter:

$$P_{q_\perp}(k) = \int \frac{d^3k'}{(2\pi)^3} \left[(1 - \mu'^2) P_{\delta\delta}(|\mathbf{k} - \mathbf{k}'|) P_{vv}(k') - \frac{(1 - \mu'^2)k'}{|\mathbf{k} - \mathbf{k}'|} P_{\delta v}(|\mathbf{k} - \mathbf{k}'|) P_{\delta v}(k')\right]$$
$$+ \int \frac{d^3k'}{(2\pi)^3} \frac{d^3k''}{(2\pi)^3} \sqrt{1 - \mu'^2} \sqrt{1 - \mu''^2} \cos(\phi' - \phi'') P_{\delta\delta vv}, \tag{6}$$

$$P_{q_\parallel}(k) = \int \frac{d^3k'}{(2\pi)^3} \left[\mu'^2 P_{\delta\delta}(|\mathbf{k} - \mathbf{k}'|) P_{vv}(k') + \frac{(k - k'\mu')\mu'}{|\mathbf{k} - \mathbf{k}'|} P_{\delta v}(|\mathbf{k} - \mathbf{k}'|) P_{\delta v}(k')\right] + \int \frac{d^3k'}{(2\pi)^3} \frac{d^3k''}{(2\pi)^3} \mu' \mu'' P_{\delta\delta vv},$$

where $\hat{\mathbf{k}} \cdot \hat{\mathbf{k}}' \equiv \mu'$, $\hat{\mathbf{k}} \cdot \hat{\mathbf{k}}'' \equiv \mu''$, and the arguments of the irreducible fourth moment $P_{\delta\delta vv}$ are the same as in Eq. (5).

*Perturbative regime.*— In the weak clustering regime, the full expressions in Eq. (6) can be simplified by using the linear relation $\tilde{\mathbf{v}} = \hat{\mathbf{k}}(\dot{a}f/k)\tilde{\delta}$, giving $P_{vv}^{(1)} = (f\dot{a}/k)^2 P_{\delta\delta}^{(1)}$ and $P_{\delta v}^{(1)} = (f\dot{a}/k) P_{\delta\delta}^{(1)}$, where $P_{\delta\delta}^{(1)}(k)$ is the power spectrum of the linear $\delta$, $\dot{a}$ is the expansion rate, and $f$ is the linear growth rate $f \approx \Omega_m^{4/7}(z)$. The perturbative $P_{q_\perp}$ and $P_{q_\parallel}$ are then second order:

$$P_{q_\perp}^{(2)}(k) = \dot{a}^2 f^2 \int \frac{d^3k'}{(2\pi)^3} P_{\delta\delta}^{(1)}(|\mathbf{k} - \mathbf{k}'|) P_{\delta\delta}^{(1)}(k') \frac{k(k - 2k'\mu')(1 - \mu'^2)}{k'^2(k^2 + k'^2 - 2kk'\mu')},$$
$$P_{q_\parallel}^{(2)}(k) = \dot{a}^2 f^2 \int \frac{d^3k'}{(2\pi)^3} P_{\delta\delta}^{(1)}(|\mathbf{k} - \mathbf{k}'|) P_{\delta\delta}^{(1)}(k') \frac{k\mu'(k\mu' - 2k'\mu'^2 + k')}{k'^2(k^2 + k'^2 - 2kk'\mu')}. \tag{7}$$

Note that $\frac{1}{2} P_{q_\perp}^{(2)}$ here is the much studied OV term [5–11].

*Nonlinear regime.*— Our main interest in this paper is the high-$k$, nonlinear behavior of $P_q$ and its implications for the kinetic SZ effect. We obtain this by noting that the dominant contributions to the integrals in Eq. (6) are for $k'$ near the peak of $P_{\delta\delta}$ or $P_{vv}$. For high $k$ beyond the peak, we can drop terms of $\mathcal{O}(k'/k)$ and obtain

$$P_{q_\perp}(k) = \frac{2}{3} \int \frac{d^3k'}{(2\pi)^3} P_{\delta\delta}(k) P_{vv}(k') = 2P_{q_\parallel}(k). \tag{8}$$

The contribution from $k' \approx k$ is proportional to $P_{vv}(k)$ or $P_{v\delta}(k)$, which falls off much faster at large $k$. The fourth moment term is also negligible at all scales (see below). The result in Eq. (8) does not assume vanishing velocity-density cross correlation, but only that $P_{\delta\delta} P_{vv}$ and $P_{\delta v} P_{\delta v}$ are of similar order. The high-$k$ result depends on the velocity power spectrum only through the integral $\int d^3k' P_{vv}(k')/(2\pi)^3 = \langle v^2 \rangle$, the volume-averaged velocity dispersion, which is less sensitive to high-$k$ nonlinear effect and appears only in the overall amplitude of $P_{q_\perp}$. Nonlinear contributions to the kinetic SZ effect therefore come mostly from $P_{\delta\delta}(k)$.

We compute the kinetic SZ power spectrum using the nonlinear halo model for $P_{\delta\delta}$ [15]. In this model, mass is assumed to be distributed in a collection of spherically symmetric halos with density profile $\rho(r)/\bar{\rho} = Au(r/r_s)$, where $u(x)$ is a specified function, and the density amplitude $A$ and scale $r_s$ are functions of halo mass. Moments of density are then superpositions over halo masses with given mass function $dn/dM$ and halo-halo correlations. Statistics on small scales are dominated by contributions from particles in a single halo; for example, the nonlinear mass power spectrum is

$$P_{\delta\delta}(k) = \int dM \frac{dn}{dM} [Ar_s^3 \tilde{u}(kr_s)]^2, \tag{9}$$

where $\tilde{u}(q)$ is the Fourier transform of $u(x)$. The halo model agrees well with numerical simulations for second and third moments of density and for pair velocity moments, and allows *ab initio* construction of and physical insight into a wide variety of density and velocity statistics [15]. In the halo model the contribution from the irreducible fourth moment to $P_{q_\perp}$ vanishes since $P_{\delta\delta vv}$ has no dependence on $\phi'$, $\phi''$, while its contribution to $P_{q_\parallel}$, related by the equation of continuity to $\dot\delta$, is found to be negligible on all scales, falling as a high power of $k$ for large $k$.

*Results.*—Figure 1 shows $P_{q_\parallel}$ and $\frac{1}{2}P_{q_\perp}$ for the density-weighted velocity $\tilde{q}$ as obtained by three different methods: the second-order OV expression of Eq. (7), the nonlinear expression Eq. (6) evaluated in the halo model of Eq. (9), and numerical simulations. The halo model results are computed using the universal profile $u(x) = x^{-3/2}(1 + x)^{-3/2}$ which approximates closely the dark matter halos found in high resolution $N$-body simulations [16]. A concentration parameter of $c(M, z) = r_{200}/r_c = 5(1 + z)^{-1}[M/(5.5 \times 10^{14} M_\odot)]^{-1/6}$ is used for the ratio of the halo virial and core radii. Figure 1 shows that the nonlinear contributions beyond second order in perturbation theory are important at $k > 1$ Mpc$^{-1}$ and must be included for reliable calculations of the kinetic SZ effect. We also note that the nonlinear halo model is able to reproduce closely the high-$k$ behavior of both $P_{q_\parallel}$ and $P_{q_\perp}$ in the simulation. Moreover, the simulation supports that $P_{q_\parallel} = \frac{1}{2}P_{q_\perp}$ on small scales and verifies that the contribution from the fourth moment $P_{\delta\delta vv}$ is unimportant on all scales. Figure 1 illustrates the validity and the importance of our nonlinear treatment.

Figure 2 shows our predictions for the angular power spectrum of temperature fluctuations due to the nonlinear kinetic SZ effect, compared with the second-order OV effect, the thermal SZ in the Raleigh-Jeans regime, and the primary CMB spectrum. The nonlinear results are obtained from Eq. (4) by integrating $P_{q_\perp}(k, z)$ in Eq. (6) to a reionization redshift of 7, an epoch suggested by recent detections of the Gunn-Peterson effect in quasar spectra [17]. The electrons are assumed to follow the $\beta$ profile $\rho_b(r)/\bar\rho = A[1 + (r/r_c)^2]^{-3\beta/2}$ with $\beta = 2/3$, which provides a reasonable overall fit to the hot gas in galaxies and clusters. We use a core radius $r_c = r_{200}/c$, with $c(M, z) = 15(1 + z)^{-1}[M/(1.8 \times 10^{13} h^{-5/2} M_\odot)]^{-0.1}$ derived from Rontgen Satellite clusters [18].

Figure 2 shows that our analytic kinetic SZ result agrees well with hydrodynamical simulations [13] when a lower halo mass cutoff of $\approx 10^{12} M_\odot$ is assumed in Eq. (9). (The two simulations in [13] have similar resolution with $2 \times 10^{10} M_\odot$ for one Cold Dark Matter particle and $1.2 \times 10^{11} M_\odot$ for 32 baryon particles.) The predicted $C_\ell$ at high $\ell$ generally increases if more lower mass halos are assumed to host hot baryons. Small-box simulations with ultra high resolution show that photoionization suppresses the baryon

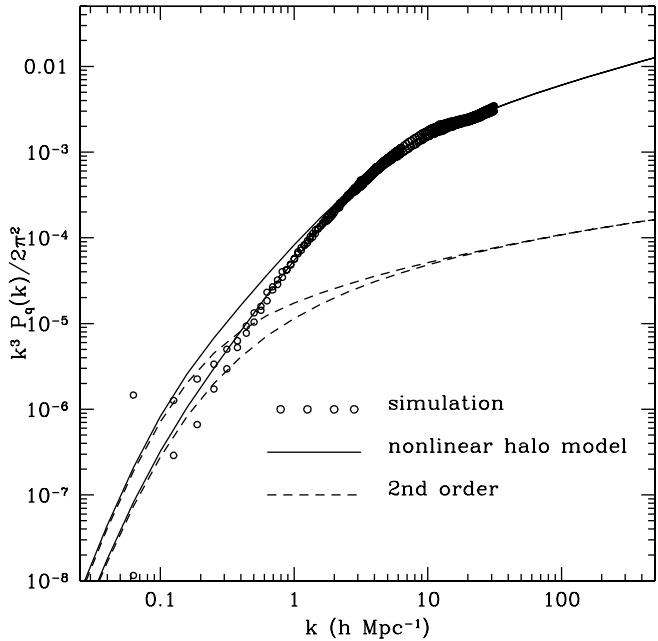

FIG. 1. 3D power spectrum of the density-weighted velocity $P_{q_\parallel}$ (upper) and $\frac{1}{2}P_{q_\perp}$ (lower). Our nonlinear analytical results (solid line) agree well with simulations (open circles), whereas the second-order OV approach (dashed line) underestimates the high-$k$ power by up to 2 orders of magnitude. The simulation results are computed directly from the particle positions and velocities in a particle-particle, particle-mesh $N$-body run with $128^3$ particles in a $(100 \text{ Mpc})^3$ comoving box and a force resolution of 30 kpc. The cosmological model is CDM with $\Omega_m = 0.3$, $\Omega_\Lambda = 0.7$, $\Omega_b = 0.05$, $h = 0.75$, and Cosmic Background Explorer normalization $\sigma_8 = 0.92$.

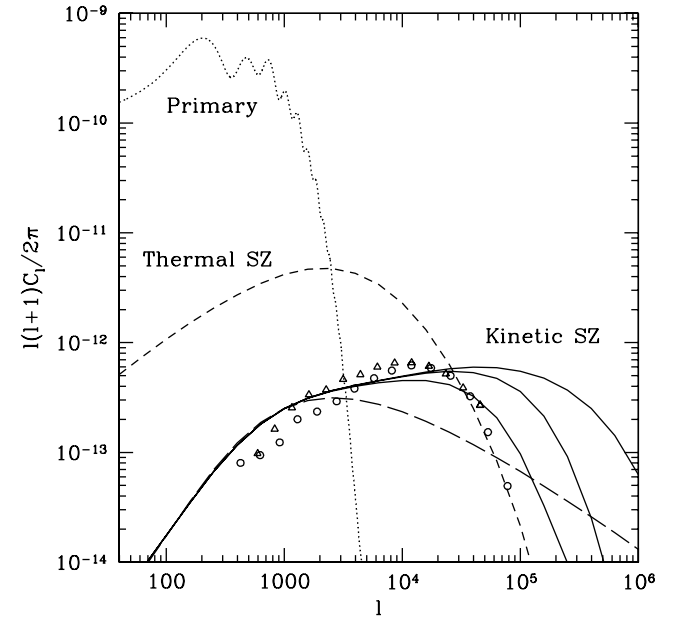

FIG. 2. Angular power spectrum for the kinetic SZ effect calculated from the nonlinear halo model (solid lines) for three lower halo mass cutoffs: $10^{12}$, $10^{11} M_\odot$, and a redshift dependent mass from [19] (bottom up). The reionization redshift is 7. For comparison we show the kinetic SZ results from the second-order perturbation theory (long-dashed line) and from the hydrodynamical simulations of Springel *et al.* (circles) and da Silva *et al.* (triangles) [13]. Also shown are the thermal SZ spectrum in the halo model (short-dashed line) and the primary CMB spectrum (dotted line). The cosmology is the same as in Fig. 1 and is very similar to that used in the simulations.

211301-3                         211301-3

fraction in low mass halos below $\sim 10^9 M_\odot$ at $z \sim 7$ to $\sim 10^{11} M_\odot$ at present [19]. The resulting kinetic SZ effect (top solid curve) based on this physical model for the minimal halo mass predicts a higher $C_\ell$ at $\ell > 40\,000$ than in the SZ simulations.

Density profiles with steeper inner slopes such as the universal dark matter profile also lead to higher $C_\ell$, but hot baryons in clusters generally follow the shallower $\beta$ profile assumed here [18]. A model of inhomogeneous reionization in a distribution of bubbles [20] shares many features of the halo model but differs in details such as halo profile shapes and in the end leads to predictions about a factor of 10 smaller. Kinetic SZ predictions from a superposition of clusters are also presented in [21]. Their model does not contain halo correlations so it does not reproduce the OV result at small $\ell$. They find the lower halo mass cutoff to have little effect on $C_\ell$, which is consistent with our results up to $\ell \sim 10^4$, but Fig. 2 here shows the strong effect at higher $\ell$.

*Discussion.*—We have derived analytical expressions for the nonlinear kinetic SZ effect on the CMB and computed the anisotropy power spectrum $C_\ell$ from the nonlinear power spectrum $P_q$ of the density-weighted velocity $\mathbf{q} = (1 + \delta)\mathbf{v}$ using the recently developed halo model of nonlinear clustering. All scales where the kinetic SZ effect is not overwhelmed by the primary CMB anisotropy are in the regime of strongly nonlinear clustering, where the second-order OV calculation underestimates the result and a fully nonlinear formulation is essential.

We have studied both the transverse and longitudinal components of $\mathbf{q}$ and found that $P_{q_\perp} \approx 2 P_{q_\parallel}$ at high $k$. Although only $\tilde{q}_\perp$ contributes to the kinetic SZ effect, we note that $\tilde{q}_\parallel$ is directly proportional to the rate of growth of the density field, so a realistic nonlinear model for $\dot{\delta}$ can, in principle, be used to compute the nonlinear kinetic SZ effect. We find both analytically and in simulations that the density-velocity fourth moment does not make an appreciable contribution to either $P_{q_\perp}$ or $P_{q_\parallel}$. In the halo model, it is easy to see that in the kinetic SZ effect contributions entering and exiting a single, spherically symmetric halo precisely cancel; the observed effect depends on the additional motions of halo centers. Contributions from rotational flow velocities or nonspherical halos may change geometric factors and alter these conclusions in detail, but the agreement with simulations in Fig. 1 shows that the halo model has incorporated the main features.

An earlier work [11] to include effects of nonlinearity started with the second-order OV expression, derived using perturbation theory (7), replaced the density power spectrum with a nonlinear form; left the linear velocity power unchanged but filtered on a scale $k_F$, and argued that density-velocity cross correlations are unimportant. The OV expression actually includes contributions from density-velocity cross correlations, which with the linear density-velocity relation enter at the same order. At large $k$, however, the contribution from the cross-correlation term is negligible due to the geometrical factor, as we found in obtaining Eq. (8). Since the perturbative expression also reduces to (8), it, by chance, produces a similar result; but our result shows that the full effect depends on the nonlinear velocity power spectrum $P_{vv}$, or in the large-$k$ or large-$\ell$ regimes through its integral over all $k$, the full velocity dispersion. We have also examined the fourth velocity-density moment $P_{\delta\delta vv}$ in the full expression (6) and demonstrated that it makes negligible contributions to the kinetic SZ effect.

We have enjoyed discussions with E. Bertschinger, B. Jain, U. Pen, M. Tegmark, and M. White. C.-P. M. acknowledges support from the Alfred P. Sloan Foundation, a Cottrell Scholars Award from the Research Corporation, Penn Research Foundation Awards, and NSF grant AST 99-73461.